\documentstyle[aps,prc,preprint,tighten,epsf]{revtex}

\def\beq{\begin{equation}}
\def\eeq{\end{equation}}
\def\bea{\begin{eqnarray}}
\def\eea{\end{eqnarray}} 

\def\eqlab#1{\label{eq:#1}}
\def\figlab#1{\label{fig:#1}}

\def\eref#1{(\ref{eq:#1})}
\def\Eqref#1{Eq.~(\ref{eq:#1})}
\def\Figref#1{Fig.~\ref{fig:#1}}
\def\figref#1{\ref{fig:#1}}

\def\al{\alpha}
\def\be{\beta}
 
\def\de{\delta} \def\De{\Delta}
\def\veps{\varepsilon}  

\def\la{\lambda}

\def\w{\omega}

\def\dd{d}

\def\ie{{i.e.}}

\def\BK#1#2{{\it #1}, #2}         
\def\CF#1#2#3#4{#1 {\bf #2}, #4 (#3)}  

\def\ann {Ann.~Phys.~(NY)}

\def\ncim {Nuovo~Cim.}
\def\np {Nucl.~Phys.}
\def\prev {Phys. Rev.}
\def\prc {Phys. Rev.~C}
\def\prd {Phys. Rev.~D}
\def\plett {Phys.~Lett.}
\def\plb {Phys.~Lett.}

\def\rq{q}

\def\arctg{{\rm arctg}}
\def\re{\mbox{Re}}

\def\3d{3-D}

\begin{document}
\title{Relativistic quasipotential equations with
$u$-channel exchange interactions}

\author{V.~Pascalutsa$^{1,2}$ and J.~A.~Tjon$^{1}$}

\address{\small\it $^{1}$ Institute for Theoretical Physics, 
University of Utrecht, Princetonplein 5, \\
3584 CC Utrecht, The Netherlands}
\address{\small\it $^{2}$ Dutch National Institute for Nuclear and High
Energy Physics (NIKHEF), \\
P.O. Box 41882, 1009 DB Amsterdam, The Netherlands\thanks{Present address.}}

\date{\today}
\maketitle

\begin{abstract}

Various quasipotential two-body scattering  
equations are studied at the one-loop level
for the case of $t$- and $u$-channel exchange potentials.
We find that the quasipotential equations devised to satisfy
the one-body limit for the $t$-channel exchange potential
can be in large disagreement with the field-theoretical 
prediction in the case of $u$-channel exchange interactions.
Within the spectator model, the description of the $u$-channel
case improves if another choice of the spectator particle is made.
Since the appropriate choice of the spectator depends strongly
on the type of interaction used, one faces a problem when both
types of interaction are contained in the potential. Equal-time
formulations are presented, which, in the light-heavy particle
system corresponding to the mass situation of the $\pi N$ system,
approximate in a reasonable way the field-theoretical result for
both types of interactions.

\end{abstract}

\section{Introduction}

In the theoretical studies of dynamics of hadronic systems
special relativity often needs to be accounted for.
Although the framework of relativistic
quantum field theory (QFT) is believed to be 
most consistent and suitable for this purpose,
it cannot be readily applied to the strongly interacting few-particle
systems without making drastic approximations.
The relativistic quasipotential (QP) equations 
\cite{Sal52,BSLT1,BSLT2,BSLT3,Grs69,Tod73,Cj88,MaW87,tlh},
present  such an approximation scheme which was extensively
applied to the description
of light nuclei, meson-nucleon, and light-quark systems.

These equations can usually be obtained from the
Bethe-Salpeter equation by truncating the kernel
and simplifying its singularity structure but keeping
the Lorentz covariance intact.
Since an infinite number of different equations can, in principle,
be derived in this way, it is desirable to establish in addition
to the requirement of Lorentz covariance other criteria,
which would constrain the choice.
For instance, an important property one would like to have for a 
relativistic two-body equation
is the {\it correct one-body limit}, meaning that in the limit
when one of the particles becomes infinitely heavy the equation
must reduce to the corresponding equation of motion of the light 
particle (e.g., the Klein-Gordon equation) in an external potential. 
Some of the first equations of this type were suggested
by Gross \cite{Grs69} and by Todorov \cite{Tod73}, and later on
other QP equations were adjusted to 
satisfy this limit \cite{Cj88,MaW87}. 

In all these investigations of the one-body limit the use of the 
$t$-channel type of potential is implicitly assumed. The quality of
the quasipotential approximations has been studied for some of
these prescriptions \cite{FT,RAP} in the equal-mass scattering case.
It was in particular found that the differences with the field theory
predictions  are moderate in magnitude and that 
the same energy dependence of the scattering amplitude can be recovered
by changing slightly  the coupling parameters. 

The aim of this paper is to examine the situation for the
$u$-channel potential. The motivation comes from the
study of the $\pi N$ scattering equations with
potentials containing both meson and baryon exchanges.
In that case the various predictions differ considerably.
In this connection it is relevant to recall that applying
the spectator equation to the $\pi N$ system Gross and Surya\cite{GrS93}
have argued that the light particle (the pion) must 
be taken as the spectator, in contrast to the original spectator
equation which demands the heavy particle on mass 
shell \cite{Grs69,Grs94}. 
Studying the box and the crossed-box graphs at threshold,
they conjecture that ``the essential difference
is the mass of the exchanged particle.'' Here we shall analyze the graphs
for more general situations, and find that the argument should be related 
to the type of the potential, rather than
to the mass of the exchange particle.

In the next section we evaluate the box and crossed-box graph
contributions for the case of $t$- and $u$- channel forces and
describe various quasipotential approximations used in the actual
studies of the $\pi N$ system. In Sec.\ III the 
field theory box graph results are compared
with the quasipotential approximations to these graphs.
We in particular do this for the situation of the unequal 
mass scattering case, corresponding to  masses of the 
$\pi N$ system.
It is found that equal-time approximations can be formulated, which
are reasonable for both types of exchanges. In Sec.\ IV
direct comparison is made at the one-loop level of the 
phase-shift predictions as obtained from the  $[1,1]$ Pad\'e
approximant to the  first two terms of the Born series of the
$K$ matrix. Some concluding remarks are made in the last section.

\section{One-loop contributions}

\subsection{Field-theoretical box graphs}

In this paper we for simplicity confine ourselves
to the case of scalar ``nucleons''. The distinct differences between
the various prescriptions can already be seen by studying this 
simplified case.
We consider the two types of the potential in \Figref{boxpotf}:
(a) $t$-exchange potential, (b) $u$-exchange potential.
Substituting these into the scattering equation, \Figref{bsef}, and
iterating once, we obtain the box graphs depicted in Figs.\ \figref{boxboxf}(a)
and \figref{boxboxf}(b), respectively. In QFT one, in 
addition, has at this level
the corresponding crossed-box graphs. 
Let us refer to the dashed line particle as  
the pion, being the light particle,
and the solid line as to the nucleon, being the heavy particle, 
(even though all the particles are scalar in our consideration) 
with corresponding masses $m_\pi$ and $m_N$.  
Obviously, the box and crossed-box graphs for both 
situations can generally be represented by \Figref{boxes}, 
where for the case (a) $m_a=m_\pi$, $m_b=m_N$, while
for the case (b) $m_a=m_N$, $m_b=m_\pi$.

Let us further denote $p,k$ and $p',k'$, the initial and final 
momenta of the external particles (taken on their mass shell:
$p^2={p'}^2=m_N^2$, $k^2={k'}^2=m_\pi^2$) and let $P=p+k=p'+k'$ 
be the total four-momentum. We now define the relative
momenta of the initial and final state as
\bea
 l &=& \be p - \al k,\nonumber\\
 l' &=& \be p' - \al k',
 \eea
 where $\al=p\cdot P/s\equiv\al(s)$, 
$\be=k\cdot P/s\equiv\be(s)$, see \Eqref{triangle}. 
The Mandelstam invariants are given by
\bea
s&=& (p+k)^2=P^2,\nonumber\\
t&=&(p-p')^2=(k-k')^2=(l-l')^2,\\
u&=&(p-k')^2=2m_N^2+2m_\pi^2-s-t\, .\nonumber
\eea
Note that $l_0=l_0'=0$ in the center-of-mass (c.m.) system defined
by $P=(P_0,\vec{0})$.

Defining the relative momentum of the intermediate state in the same
way, \ie, $q=\be p''-\al k''$,
where $p''$ ($k''$) is the intermediate nucleon (pion) momentum, the box
graph of \Figref{boxboxf}(a) corresponds to 
\bea
\eqlab{boxqft}
B(s,t)&=&\frac{-i}{\pi^2}\int \!\dd^4 q 
\, \frac{1}{[(q-l)^2 - \mu^2+i\veps]\,[(q-l')^2 - \mu^2+i\veps]} \nonumber \\
&\times&\frac{1}{[(\al P+q)^2 - m_b^2+i\veps]\,[(\be P-q)^2-m_a^2+i\veps]}
\eea
with $m_a=m_\pi,\, m_b=m_N$, and $\mu$ the mass of the exchanged particle.
We can write the $u$-channel box of \Figref{boxboxf}(b) also in this 
form by introducing as the momentum of  integration 
$q=\be k''-\al p''$ and taking $m_a=m_N,\, m_b=m_\pi$.

Consider now the poles of the integrand in the complex $q_0$ plane:
\bea
\eqlab{poles}
(1)&\,\,\,&q_0 =  - \sqrt{(\vec{q}-\vec{l})^2+\mu^2}+i\veps, \nonumber\\
(2)&\,\,\,&q_0 =  - \sqrt{(\vec{q}-\vec{l}')^2+\mu^2}+i\veps,\nonumber\\
(3)&\,\,\,&q_0 = -\al P_0 - \sqrt{(\al\vec{P}+\vec{q})^2+m_b^2}+i\veps,\nonumber\\
(4)&\,\,\,&q_0 = \be P_0 - \sqrt{(\be\vec{P}-\vec{q})^2+m_a^2}+i\veps, \nonumber\\
(5)&\,\,\,&q_0 = \sqrt{(\vec{q}-\vec{l})^2+\mu^2}-i\veps, \\
(6)&\,\,\,&q_0 = \sqrt{(\vec{q}-\vec{l}')^2+\mu^2}-i\veps,\nonumber\\
(7)&\,\,\,&q_0 = -\al P_0 + \sqrt{(\al\vec{P}+\vec{q})^2+m_b^2}-i\veps,\nonumber\\
(8)&\,\,\,&q_0 = \be P_0 + \sqrt{(\be\vec{P}-\vec{q})^2+m_a^2}-i\veps.\nonumber
\eea
To perform the integration in \Eqref{boxqft} over the relative energy 
variable $q_0$ we apply the Wick rotation:
$q_0\rightarrow iq_0$. The rotation has to be made in such a 
way that the poles which can cross the ${\rm Im}\, q_0$ axis, 
when the spatial integration variable varies,
are avoided. Provided that no pinching of 
singularities occurs such a rotation can be carried out. 
For the crossed box of Fig. 3(b) this indeed does not occur. For the
direct box, poles 4 and 7 do pinch when we are in the scattering 
region, and hence in that case the Wick rotated integration has to be 
deformed to a contour\cite{LWT67} as shown in \Figref{wickf}.
The integral over $q_0$ is thus equal to the singularity-free 
integration along the imaginary axis plus the residues of the 
two poles, \ie, in the c.m.\ system we have
\bea
\eqlab{integrate}
B(s,t)&=&\frac{1}{\pi^2}
\int\!\dd\Omega\left\{ \int\limits_0^{\infty}\!\dd\rq\,\rq^2
\int\limits_{-\infty}^{\infty} \!\dd q_0 
\frac{1}{[(\al P_0+iq_0)^2- \w_b^2]\,
[(\be P_0-iq_0)^2-\w_a^2]} \right. \nonumber\\
&\times & \frac{1}{[-q_0^2-(\vec{q}-\vec{l})^2 - \mu^2]\,
[-q_0^2-(\vec{q}-\vec{l}')^2 - \mu^2]} -  \pi\! 
\int\limits_0^{\hat{q}}\!\dd\rq\,\rq^2\, \left[
\frac{1}{\w_a\,[(P_0-\w_a)^2-\w_b^2]} \right. \nonumber \\
&\times &
\frac{1}{[(\be P_0-\w_a)^2-(\vec{q}-\vec{l})^2 - \mu^2]\,
[(\be P_0-\w_a)^2-(\vec{q}-\vec{l}')^2 - \mu^2]} \\ 
&+&\left.\left. \frac{1}{\w_b\,[(P_0-\w_b)^2-\w_a^2]\,
[(\al P_0-\w_b)^2-(\vec{q}-\vec{l})^2 - \mu^2]\,
[(\al P_0-\w_b)^2-(\vec{q}-\vec{l}')^2 - \mu^2]} \right]\right\},\nonumber
\eea    
where $\w_i=(\rq^2+m_i^2)^{1/2}$, $\hat{q}^2=\la(s)/s$,
and $\la$ is the triangle function defined in \Eqref{triangle}.
We have used \Eqref{integrate} to evaluate numerically the box and
crossed-box contributions.

\subsection{Quasipotential approximations}

Let us now define the box graphs obtained within various 
QP formalisms.
In applying the {\it spectator} prescription, only one of the poles 
in $q_0$ is taken into account. For example, if particle $m_a$ is
the spectator, only pole 4 of \Eqref{poles} is taken, and one has
\bea
B_{\rm spect}&=& - \frac{1}{\pi}
 \int\!\dd^3 q \,
\frac{1}{\w_a\,[(P_0-\w_a)^2-\w_b^2]\,
[(\be P_0-\w_a)^2-(\vec{q}-\vec{l})^2 - \mu^2]}\nonumber\\
&\times&\frac{1}{[(\be P_0-\w_a)^2-(\vec{q}-\vec{l}')^2 - \mu^2]}.
\eea 

In the {\it equal-time} (ET) approximation, the retardation in the 
exchanged particle propagators is neglected. It means that, in the
c.m.\ system, the relative energy 
is set to {\it zero} in the propagators of
particle $\mu$, while the poles of $m_a$ and $m_b$ are treated exactly, 
\ie, for case (a):
\beq
\eqlab{etbox-a}
B_{\rm ET} (s,t)=\frac{1}{\pi}\int \!\dd^3 q
\frac{1}{[-(\vec{q}-\vec{l})^2 - \mu^2]\,[-(\vec{q}-\vec{l}')^2 - \mu^2]}\,
G_{\rm ET} (\rq^2,s),
\eeq
while for the case (b):
\beq
\eqlab{etbox-b}
B_{\rm ET}(s,t)=\frac{1}{\pi}\!\int \!\dd^3 q
\frac{G_{\rm ET} (\rq^2,s)}{\left[\frac{(m_N^2-m_\pi^2)^2}{s}
-(\vec{q}-\vec{l})^2 - \mu^2\right]\,\left[\frac{(m_N^2-m_\pi^2)^2}{s}
-(\vec{q}-\vec{l}')^2 - \mu^2\right]},
\eeq
where 
\bea
G_{\rm ET} (\rq^2,s)&=&\frac{-i}{\pi}\int\!\dd q_0 
\frac{1}{[(\al P+q)^2 - m_N^2+i\veps]\,[(\be P-q)^2-m_\pi^2+i\veps]} \nonumber\\
&=&\frac{1}{\la(s)/s-\rq^2+i\veps}\left(\frac{\al(s)}{\w_N}+
\frac{\be(s)}{\w_\pi}\right).
\eea
It should be remarked that for the scalar case being studied
the above-defined ET formalism is equivalent to the one of
Salpeter \cite{Sal52}.
 
In the {\it symmetrized equal-time} prescription of Mandelzweig and Wallace
\cite{MaW87}, the contribution from the forward-scattering crossed box is
approximately included by modifying the two-particle Green function
\bea
G_{sym\rm ET} (s,t)&=&G_{\rm ET} (\rq^2,s)+
G_{\rm ET} (\rq^2,2m_\pi^2+2m_N^2-s).
\eea

We would like now to consider one more prescription, motivated by the fact
that the $u$-exchange case of the ET approximation suffers from the
exchanged particle singularities when condition 
\beq
\eqlab{cond}
\mu^2 - (m_N^2-m_\pi^2)^2 (s-\hat{q}^2)/s^2 \leq 0
\eeq 
is satisfied.
To avoid this we may write down both cases
in the form of \Eqref{boxqft} and then make the approximation $q_0=0$
in there. For the $t$-exchange case this is just the usual 
ET approximation, while for the $u$ exchange  this implies 
an energy transfer equal to $(\be-\al)P_0$. 
In the latter case, we therefore refer to this prescription as to 
{\it constant energy-transfer} (CET) approximation. 
We should note that, in this approximation, 
both cases (a) and (b) become fully 
equivalent ($t$-exchange ET equal to $u$-exchange CET),
since the two-particle propagator $G_{\rm ET}$ is obviously symmetric
with respect to the interchange of $m_a$ and $m_b$. In analogy to
the symmetrized ET, the {\it symmetrized CET} approximation can be defined
and will be studied as well. 

\section{Comparison in the forward direction}

We have calculated the field theoretical scalar box and crossed-box graphs 
in (3+1) dimension both numerically by integrating
\Eqref{integrate}, and analytically for the forward
scattering with the explicit expressions  given in the Appendix.
For the numerical integrations we used the Gaussian quadrature method. We
observed that in order to obtain a good numerical stability, especially near
threshold, one needs to take
a rather large number of Gaussian points for the $q_0$ integration
(we have taken 320 points). On the other hand, 64 points for q, and 
8 points for each of the angular integrations is sufficient. We have 
checked that our numerical calculation agrees, for $t=0$, with 
the analytical expressions of the Appendix and for $t\neq0$, with the
code developed by Veltman \cite{PaV79}.
 
Using the equations described in the previous section
we have also determined numerically the QP box graphs.
We confine ourself in this section to the forward direction, i.e., $t=0$.
Similar results are found for $t\ne 0$.
In the following we consider the real part of the one-loop
contributions. The imaginary part is essentially determined from the
two-particle unitarity condition.
As a typical example we show in \Figref{ex1} the dependence of the box
on the exchange mass $\mu$ for the case that the heavy particle
is much heavier than the light one.
We have taken $m_N=1$, $m_\pi =0.01$\footnote{Note that we 
multiply the results by $\mu^3 m_N$  in order to obtain
reasonable values for various limiting values of $\mu$ and $m_N$,
since, for instance, at threshold we have for the $t$-exchange case:
$$
\lim_{\mu\rightarrow 0} B [(m_N\pm m_\pi)^2,0] = 
 \frac{\pi}{2 \mu^3 m_N} 
\frac{1\pm 1}{1\pm \frac{m_\pi}{m_N}}.
$$ }
and the energy is fixed somewhat
above threshold, $\sqrt{s}=1.1 m_N$.

From the figure we see that for the $t$-exchange potential 
the one-body limit is achieved\footnote{The
 proof of the correct one-body limit given at the one-loop
level can usually be extended for the whole equation, see, e.g.,
\cite{Grs94,SiT93}. In our discussion we shall therefore assume that 
the one-body limit is 
satisfied in a given QP formulation if, in the limit, 
the QP box graph becomes equal to the sum of the 
field-theoretical box and crossed-box graph.}
in the symmetrized ET formulation {\it independently} of the
mass of the exchanged particle. The nucleon spectator approximation
clearly deviates from this limit for large $\mu$. However,
we also can see that the pion spectator gives an even worse prediction.
In the $u$-exchange case, both the nucleon spectator 
and symmetrized
ET disagree substantially with the QFT result (the spectator 
calculation is an order
of magnitude larger and hence beyond the scale of the figure), 
while the pion spectator prediction is in good agreement.
Based on these observations we can, in particular,
conclude that the difference
between the $NN$ situation and the $\pi N$ situation encountered
by Gross and Surya \cite{GrS93} appears
due to the different {\it type of potential}.

We were unable to study the $u$-channel exchange
for smaller $\mu$, because of the occurrence of the exchange
particle singularities, due to condition \eref{cond}.  
Only the constant energy-transfer
approximation (CET) can, in principle, be discussed in the 
whole mass region. In \Figref{excet1}
we compare the CET prescription with the
exact result for the $u$-channel situation. We see that for large 
$\mu$ the QP and exact calculation converge to the same answer.
Recall that CET for the $u$ exchange is just equal 
to the ET for the $t$ exchange.
This in particular indicates that for large $\mu$ the exact result for
$t$ and $u$ exchange should be the same. For smaller $\mu$ the exact
$u$-exchange results are strongly affected by the above-mentioned
$\mu=m_N$ singularity in the separate box and crossed-box contributions.

The qualitative difference between the $t$ and $u$ exchange in the
$m_{\pi}/m_N\rightarrow 0$ limit is
transparently seen from the analytical expressions. 
For $t$-exchange we have at threshold
\bea
\eqlab{vgl12}
B[(m_N\pm m_\pi)^2,0] &=&\frac{2}{\mu^3(m_N\pm m_\pi)}
\left\{ \frac{m_N}{\sqrt{(2m_N)^2-\mu^2}}\,
\arctg\sqrt{\left(\frac{2m_N}{\mu}\right)^2-1} \right.\nonumber\\
&\pm& \left.\frac{m_\pi}{\sqrt{(2m_\pi)^2-\mu^2}}\,
\arctg\sqrt{\left(\frac{2m_\pi}{\mu}\right)^2-1}\right\} .
\eea
From  \Eqref{vgl12} we see that 
for small $m_\pi$ (or, equivalently, large $m_N$ and arbitrary $\mu$),
there is a cancellation between the box and 
crossed box leading to the following result:
\beq
B[(m_N + m_\pi)^2,0] + B[(m_N- m_\pi)^2,0] =
 \frac{4}{\mu^3 \sqrt{4m_N^2-\mu^2}}\,
\arctg\left[\sqrt{\left(\frac{2m_N}{\mu}\right)^2-1}\right].
\eeq
In contrast, 
in the case of the $u$-exchange at threshold, both the box and 
the crossed box 
{\it vanish}, for $\mu\neq m_N$. The special case $\mu=m_N $ is 
singular, yielding
$$
\lim_{m_\pi\rightarrow 0} B [(m_N\pm m_\pi)^2,0] = 
\pm\frac{1}{2 m_\pi m_N^3}.
$$
Hence only the sum of the $u$-exchange boxes vanishes.

It is still remarkable though that the pion spectator is so 
close to the exact result for the $u$-exchange case, suggesting that
the dominant pole always comes from particle $m_b$. This
can be understood by using the crossing relation between the
$t$- and $u$-channel box graphs: under the crossing the 
field-theoretic $t$-channel box and crossed box
turn into the corresponding $u$-channel graphs, while
the nucleon-spectator box turns into the pion-spectator box.

We have also studied somewhat more realistic situations, away from
the one-body limit. In \Figref{ex2}
we plot the results for $m_\pi=0.15 m_N$ (\ie, the physical pion mass),
and $\mu=m_N$. In \Figref{ex3} we have taken $m_\pi=0.15 m_N$, and
$\mu=0.82 m_N$ (the $\rho$-meson mass) for $t$ exchange, while
$\mu=1.31 m_N$ (the $\De$-isobar mass) for $u$ exchange. The corresponding 
CET calculations are presented in \Figref{excet2}.
From these figures we see that for the $t$-exchange all of the QP
prescriptions, except the pion spectator one, do reasonably well as they
have the correct energy dependence, and the small discrepancy
in the magnitude can possibly be accounted for by a 
re-adjustment of the coupling strength. 

For the $u$ exchange the ET and the pion spectator prescriptions do 
particularly well, especially for larger exchange mass and/or
larger energy. In \Figref{ex3} they both practically ``fall on top''
of the exact result. From \Figref{excet2} note that the ordinary
CET agrees in overall better with QFT than the symmetrized version.
The good agreement at large energies is again very remarkable.

Apparently, in a model where both the $t$-
and $u$-exchange potential are present, the spectator approximation
does not provide an optimal choice. Choosing the pion spectator
leads to pathological results for the iterated $t$-exchange potential. 
A similar contradiction appears in the nucleon spectator prescription
and the $u$-exchange potential. On the other hand,
the ET prescriptions, including CET and the symmetrized versions,
are preferable from this point of view. We cannot make a
definite preference among the two, although some  agreement 
between a given choice in the ET approach and the
exact results are observed in certain regimes, which
perhaps should be studied in more detail. 

It must be emphasized,
that the one-body limit situation is physically very different
for the two cases: for the $t$-exchange potential it corresponds to
the light particle moving in an external potential of the heavy 
particle, while in the  $u$-exchange case the heavy particle 
obviously does not act as a static external source, and, therefore, 
there is no correspondence to any one-body situation. 
Clearly, the possibility to have a QP approach, which
describes at the same time both cases of $t$- and $u$-exchange  
in a reasonable way is interesting.

\section{Phase-shift calculations}

We may also examine the differences among the various
prescriptions at the level of phase shift predictions.
Similar to the $NN$ scattering case{\cite{ZuT82} this can be done
by reconstructing the scattering amplitude from the driving force
and the one-loop contributions.
For this we assume the following $t$- and $u$-channel
potentials (in a $\phi^3$ theory):
\bea 
V_t (p,q) &=& \frac{g^2}{4\pi}\frac{\mu^2}{\mu^2-(p-q)^2-i\veps}, \nonumber\\
V_u (p,q;P) &=& \frac{g^2}{4\pi}\frac{\mu^2}{\mu^2-(p+q-P)^2-i\veps}, \nonumber
\eea
as the driving force in the scalar Bethe-Salpeter 
equation\footnote{In our convention
$V$ and $T$ contain an extra
factor of $1/4\pi$, hence the usual volume factor $(2\pi)^4$ is 
replaced by $4\pi^3$.}
\beq
\eqlab{bse}
 T(p',p;P)=V(p',p;P)+ i\int\! \frac{\dd^4 q}{4\pi^3}\,V(p',q;P)\,G(q;P)
\,T(q,p;P).
\eeq
Introducing the $l$th partial wave $K$ matrix
\beq
K_l=\frac{T_l}{1+i \hat{\w} T_l},
\eeq
where $\hat{\w} = \sqrt{\la(s)}/s$,
we obtain from Eq. \eref{bse} the $K$-matrix equation (omitting
external momenta):
\bea
\eqlab{kmat}
K_l &=& V_l + \frac{i}{\pi^2} \int\limits^\infty_{-\infty} \dd q_0
\,{\cal P}\! \int\limits^\infty_{0} 
\dd \rq\, \rq^2 \, V_l(q_0,\rq) G(q_0,\rq) K_l(q_0,\rq) \nonumber\\
&=& V_l + \frac{i}{\pi^2} \int\limits^\infty_{-\infty} \dd q_0
\,{\cal P}\! \int\limits^\infty_{0} 
\dd \rq\, \rq^2 \, V_l(q_0,\rq) G(q_0,\rq) V_l(q_0,\rq) + \ldots \nonumber\\
&\equiv & K^{(0)}_l + K^{(1)}_l +\ldots,
\eea
where ${\cal P}\int \dd \rq$ is the principal value integral, while
$V_l =\frac{1}{2}\int^{1}_{-1} \dd x \, V \, P_l(x)$ is the 
partial wave decomposed potential,
$x$ being the cosine of the center-of-mass scattering angle.
The second term in Eq. \eref{kmat} can immediately be written in 
terms of the field theory box graph $B$:
$$ K^{(1)}_l =-\frac{(g\mu)^4}{(4\pi)^2}\, 
(1/8\pi) \int\limits^{1}_{-1}\dd x\, \re B\, P_l(x).$$
From the $K$ matrix we can determine the phase shift 
through the relation
\beq
{\rm tan}(\delta) = \hat{\w} K_l.
\eeq

The series for $K_l$ can be summed by Pad\'e approximants to get a 
converged solution of the integral equation.
When we confine ourselves to the study of the box graphs we can carry 
out a geometric summation of the Born series, being essentially
the [1,1] approximant in the coupling  constant $g^2$, i.e.,
\beq
\eqlab{PA11}
K^{[1,1]}_l = K^{(0)}_l\,
\left(1-\frac{K^{(1)}_l}{K^{(0)}_l}\right)^{-1}.
\eeq
For not too strong coupling we expect that this is a
reasonable approximation to the solution of Eq. \eref{kmat}
\cite{GM}.
In  \Figref{phases} we  show  the phase shifts
obtained  in the various calculations of the
$K^{[1,1]}$ approximant, together with the Born approximation
$K_l=K^{(0)}_l$. 
The depicted results correspond to masses
relevant to the $\pi N$ system, while the coupling
strength,  taken to be $g^2/(4\pi)=2.0$, 
has been  adjusted such that the relative size of rescattering
effects is of the same order of magnitude as observed in
our realistic $\pi N$ calculations\cite{PT99}. 

Also are shown the results which 
include the crossed-box graph in the field-theory calculation.
This is determined by approximating the $K$ matrix as
\beq
K^{[1,1]}_l = K^{(0)}_l\,
\left(1-\frac{K^{(1)}_l+B^{X\mbox{box}}_l}{K^{(0)}_l}\right)^{-1}.
\eeq
with $B^{X\mbox{box}}_l$ being the corresponding partial wave 
reduced crossed-box diagram.
One can see that the crossed box contributions are rather small
and that they give rise to an additional attraction in the 
$S$-wave channel. One can also state that the predictions shown in 
\Figref{phases} are qualitatively similar to the results
obtained in the forward direction. 

To get a feeling 
on the convergence of the Born series for the used coupling
constant, we compare in \Figref{rescat} the change due to
including the second Born term perturbatively, i.e.,
$K_l=K^{(0)}_l + K^{(1)}_l $.
From the figure we see that the higher-order correction is
rather moderate, so that we expect that the $[1,1]$ approximant
is reasonable for the considered strength of the coupling
constant.

From the present analysis we find that the nucleon and pion 
spectator models do lead to a 
reasonable description of the phase shift for the case of 
$t$- and $u$- channel exchanges, respectively, but do not describe
both types of exchanges simultaneously. For instance, the 
nucleon-spectator model with the $u$-exchange potential,
leads to a considerably stronger attraction than would have
been predicted by the field-theory box graph contribution. 
Assuming that the $[1,1]$ Pad\'e approximant is valid,
the nucleon-spectator model predicts 
the existence of a bound state in the $S$ wave, to be
contrasted with none in the other quasipotential
prescriptions. As a consequence the predicted phase shift is
distinctly different in this case as compared to the other
quasipotential predictions. It decreases from $\pi$ at threshold
to zero with increasing energy, as can be inferred from
\Figref{phases}. 

\section{Conclusions}

We have studied here in detail various quasipotential approximations 
to the box graph and compared them with the field-theory graphs.
We have chosen the kinematics of $t=0$ to present the various
comparisons of the one-loop contributions.
Although the present study has been confined to the situation of scalar
particles, the same conclusions apply when one considers the case of fermions.
Moreover, similar results  were found when the phase shifts are studied
up to the one-loop level.  

In the large external mass ratio limit a large qualitative difference 
is observed between the situation when the potential in question has
the form of $t$- or $u$-channel particle exchange. 
The QP equations, such as the nucleon spectator \cite{Grs69,Grs94}
and the symmetrized ET \cite{MaW87},
developed to satisfy the one-body limit for the $t$-type exchange 
potential, have a poor agreement with the exact calculation if the
$u$-type exchange potential is used. The differences are in general so
large, that large  reductions  of the coupling constants will
be needed to establish reasonable agreement of the phase shifts.
Although the pion spectator 
approximation describes the $u$-exchange case better, it however fails
in the other case. Therefore, in the situation where both types of the 
potential are present, either of the spectator equations cannot be 
justified. 

It appears that in the case of the $u$-channel exchange potential 
the one-body limit cannot be viewed analogously to
the $t$-channel case, essentially because in the former case
the heavy particle is not expected to act like a source. 
Thus, in general, the one-body criterium for quasipotential equations 
should be reconsidered. Instead one can for instance demand 
that the quasipotential prescription leads 
to a good approximation of the field theory box graphs.

Analyzing the situation with, for the $\pi N$ 
system realistic parameters, we find that the ET type of prescriptions
can be fairly close to the QFT answer for both types of the potential.
We may hope that such a ET prescription may offer 
us a suitable dynamical framework  to describe the $\pi N$ system.
The above study clearly indicates that these quasipotential formulations
have indeed very nice properties to render an attractive framework for 
application to the $\pi N$ system. It can clearly treat both the 
$t$- and $u$-exchange forces in a reasonable way.
In a separate publication we report on the results of a
relativistic study of the $\pi N$ dynamics, based on hadron degrees of
freedom, including the full spin complication and employing
the equal-time quasipotential formalism\cite{PT99}.

\vspace{1cm}
\noindent
{\bf Acknowledgments}
\vspace{0.3cm}

\noindent
This work was partially financially supported by de Stichting voor
Fundamenteel Onderzoek der Materie (FOM), which is sponsored by the
Nederlandse Organisatie voor Wetenschappelijk Onderzoek (NWO).

\appendix
\section*{Forward-scattering box graphs}
For $t=0$, using the Feynman parameter trick, we can 
rewrite the expression for the box \Eqref{boxqft} as follows:
\bea
B(s,0) &=& \int\limits_0^1 \dd x\,\dd y\,\dd z\,
\frac{x\, \de(1-x-y-z)}{h^2(x,y,z)} 
=\int\limits_0^1 \dd x\,\dd y\,
\frac{x(1-x)}{h^2[1-x,xy,x(1-y)]}, \nonumber \\
&& h(x,y,z):= \mu^2x + m_a^2 y + m_b^2 z - m_\pi^2 xy -m_N^2 xz-syz. 
\eea
Integrations yield the following result:
\begin{itemize}
\item[(a)] $t$ exchange,  $m_a=m_\pi,\, m_b=m_N$:
\bea
B(s,0) &=& \frac{1}{N_{\rm a} (s)}\left\{ 2\,\sqrt{-\la(s)}\,
\arctg\left[\frac{s/ \sqrt{-\la(s)}}{1-s^2\al(s)\,\be(s)/ \la(s)}\right]
 \right. \nonumber\\
&-&\frac{s+m_N^2-m_\pi^2}{\sqrt{(2m_N/\mu)^2-1}}\,
\arctg\left[\sqrt{(2m_N/\mu)^2-1}\right] \nonumber \\
&-&\left. \frac{s+m_\pi^2-m_N^2}{\sqrt{(2m_\pi/\mu)^2-1}}\,
\arctg\left[\sqrt{(2m_\pi/\mu)^2-1}\right] \right\}, 
\eea
with
$N_{\rm a} (s)=\mu^2 [4\la(s)-s \mu^2]$.
\item[(b)] $u$ exchange, $m_a=m_N,\, m_b=m_\pi$:
\bea
&&B(s,0) = \frac{2}{N_{\rm b} (s)} \sqrt{-\la(s)}\,
\arctg\left[\frac{s/ \sqrt{-\la(s)}}{1-s^2\al(s)\,\be(s)/ \la(s)}\right]
 \nonumber\\
&&+\frac{1}{2N_{\rm b} (s)\,\sqrt{-\la(\mu^2)}}\left\{ 
2 [(m_N^2-m_\pi^2)^2-\mu^2 s]\,
\arctg\left[\frac{m_N^2+m_\pi^2-\mu^2}{2\sqrt{-\la(\mu^2)}}\right]
\right.\nonumber\\
&&+ [ (m_N^2-m_\pi^2)^2-s \mu^2 + (m_N^2-m_\pi^2)(\mu^2-u) ]\,
\arctg\left[\frac{\mu^2 \al(\mu^2)}{\sqrt{-\la(\mu^2)}}\right]  \nonumber\\
&&+\left. [ (m_N^2-m_\pi^2)^2-s \mu^2 - (m_N^2-m_\pi^2)(\mu^2-u) ]\,
\arctg\left[\frac{\mu^2 \be(\mu^2)}{\sqrt{-\la(\mu^2)}}\right]
\, \right\}, 
\eea
with $u=2m_N^2+2m_\pi^2-s$, and $ N_{\rm b} (s)
=(\mu^2-u)\,[(m_N^2-m_\pi^2)^2-s \mu^2]$.
\end{itemize}

Here functions $\la$, $\al$ and $\be$ are defined as
\bea
\eqlab{triangle}
\la(x)&=&[x-(m_N+m_\pi)^2]\,[x-(m_N-m_\pi)^2]/4, \nonumber\\
\al(x)&=& (x+m_N^2-m_\pi^2)/2x, \\
\be(x)&=&(x-m_N^2+m_\pi^2)/2x. 
\eea
Note that the crossed-box graph is given simply  by $B(u,0)$.

Let us also quote the threshold value for the case when all the
masses are equal, since some care is required in this calculation. 
We have for $m_N=\mu$
\bea
\lim_{m_\pi\rightarrow \mu} B[ (\mu+ m_\pi)^2,0] &=&
\frac{2\pi}{3\sqrt{3}}\,\mu^{-4} \approx 1.2092 \,\mu^{-4}, \nonumber\\
\lim_{m_\pi\rightarrow \mu} B[ (\mu- m_\pi)^2,0] &=&
 \frac{2}{3}\left(1 - \frac{\pi}{3\sqrt{3}}\right)\,\mu^{-4}
\approx 0.2636 \,\mu^{-4}.
\eea

\bibliographystyle{prsty}

\newpage
\vspace{3cm}

\begin{figure}[h]
\begin{center}
\epsffile{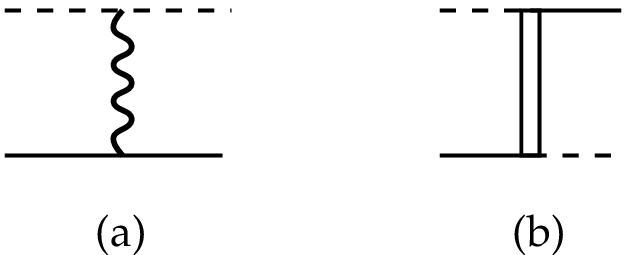}
\end{center}
\caption{ The $t$-channel (a) and $u$-channel (b) exchange potentials.}
\figlab{boxpotf}
\end{figure}

\vspace{3cm}

\begin{figure}[h]
\epsffile{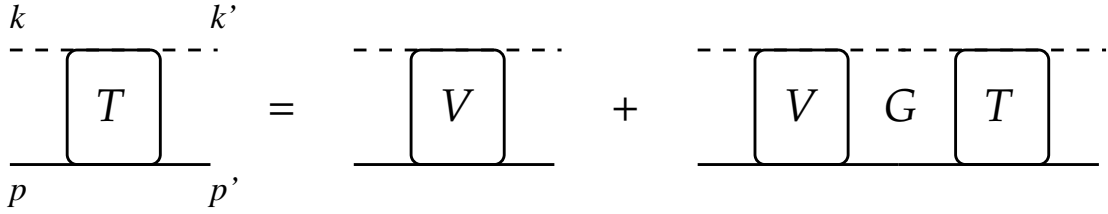}
\caption{ Diagrammatic form of a relativistic two-body scattering  
equation.}
\figlab{bsef}
\end{figure}

\vspace{3cm}

\begin{figure}[h]
\begin{center}
\epsffile{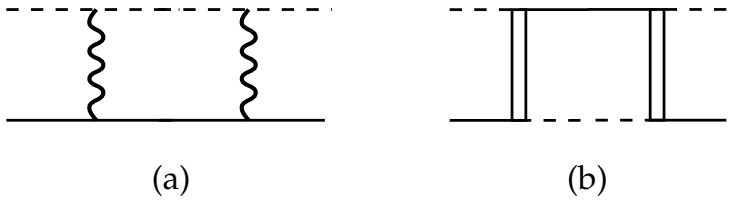}
\end{center}
\caption{ The box graphs obtained by iterating once the
potentials of \Figref{boxpotf}.}
\figlab{boxboxf}
\end{figure}

\vspace{3cm}

\begin{figure}[h]
\begin{center}
\epsffile{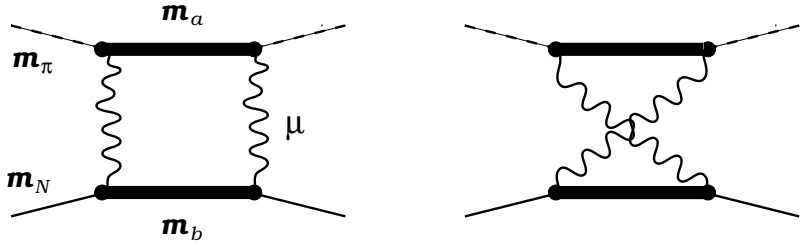}
\end{center}
\caption{ The box and crossed-box graphs with masses $m_a$ and
$m_b$ in the intermediate state.}
\figlab{boxes}
\end{figure}

\vspace{3cm}

\begin{figure}[h]
\begin{center}
\epsffile{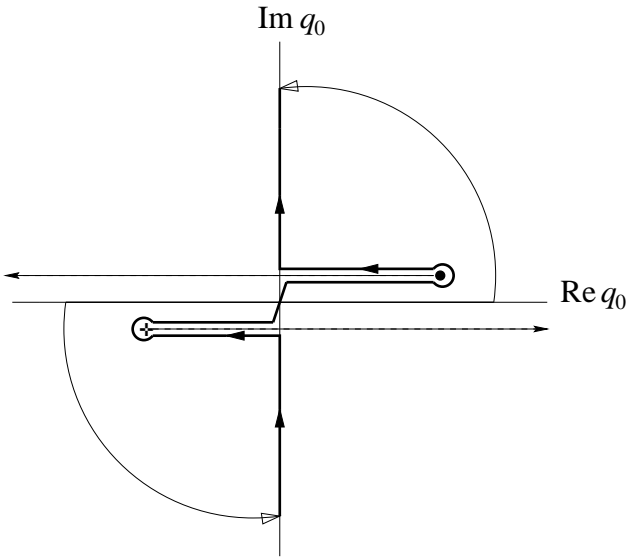}
\end{center}
\caption{ Wick rotation and the resulting integration path (bold line)
in the complex $q_0$ plane. 
The situation is shown where the two poles 4 and 7 from
\Eqref{poles} have pinched and crossed the imaginary $q_0$-axis.
}
\figlab{wickf}
\end{figure}

\vspace{3cm}

\epsfxsize=14.7cm
\begin{figure}[h]
\begin{center}
\epsffile{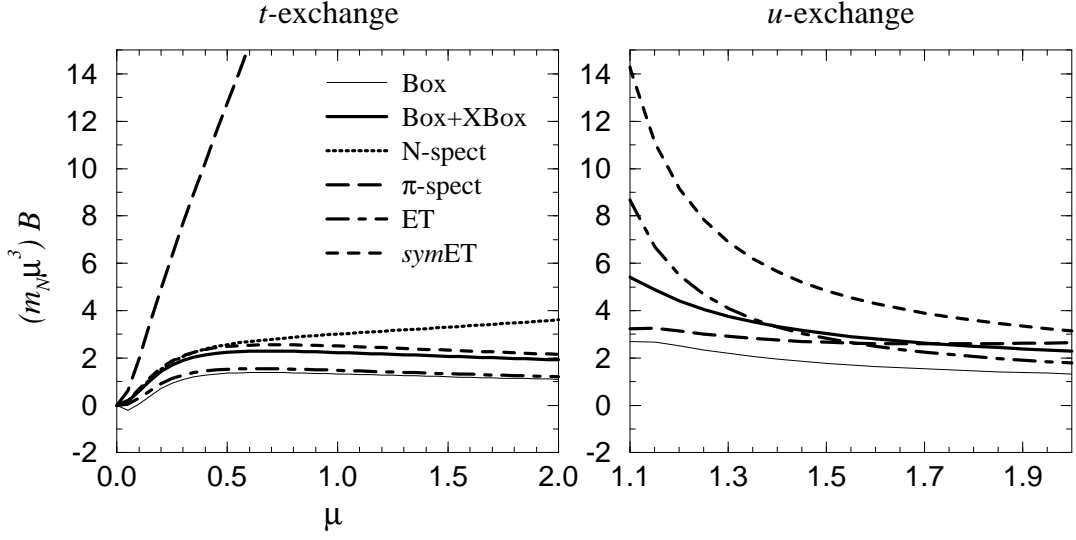}
\end{center}
\caption{ The box and various QP approximations for
$m_N=1,\,m_\pi=0.01,\, \sqrt{s}=1.1$ and $t=0$ 
as a function of the mass $\mu$ of the exchanged particle. 
In the left and right panels are shown the results corresponding
to the graphs (3a) and (3b) respectively.
Also are shown the results when the crossed-box is added to the
box contribution.
}
\figlab{ex1}
\end{figure}

\vspace{3cm}

\epsfxsize=8.3cm
\begin{figure}[h]
\begin{center}
\epsffile{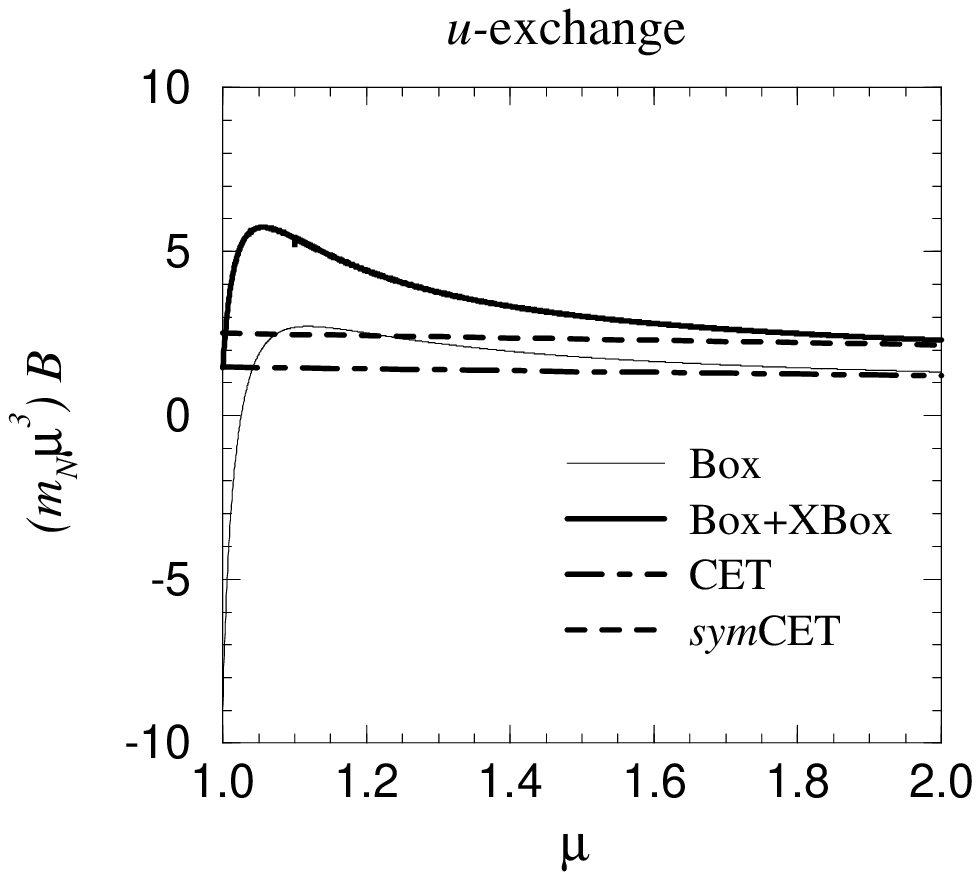}
\end{center}
\caption{ The QFT and CET results for the same set of parameters
as in Fig. 6 for the case of the graph (3b). }
\figlab{excet1}
\end{figure} 

\vspace{3cm}

\epsfxsize=14.7cm
\begin{figure}[h]
\begin{center}
\epsffile{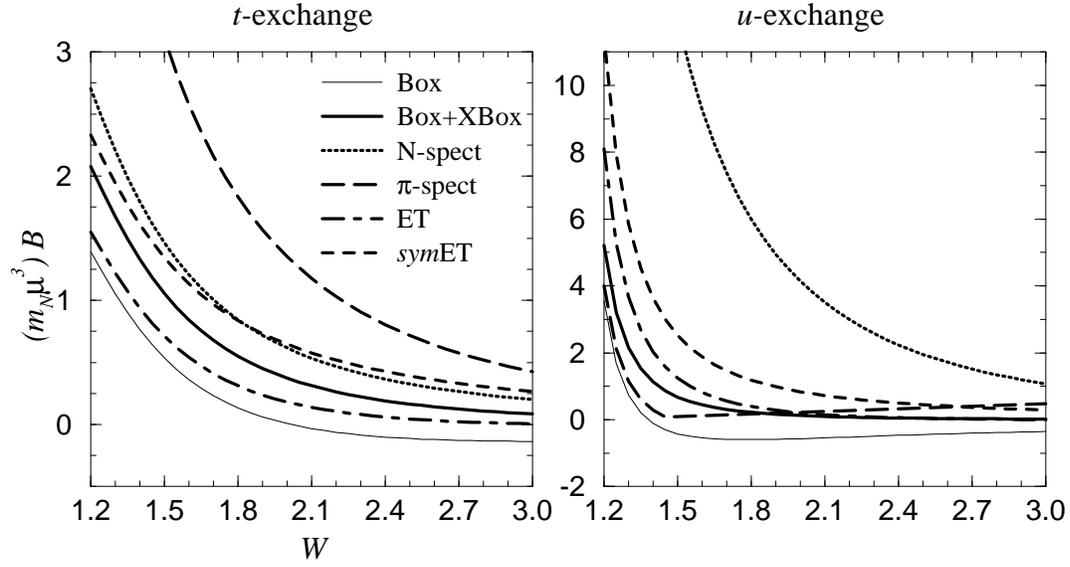}
\end{center}
\caption{  The same as Fig. 6, but 
for $m_N=\mu=1,\,m_\pi=0.15$ and $t=0$, as a function
of the energy $W=\sqrt{s}$.}
\figlab{ex2}
\end{figure}

\vspace{3cm}

\epsfxsize=14.7cm
\begin{figure}[h]
\begin{center}
\epsffile{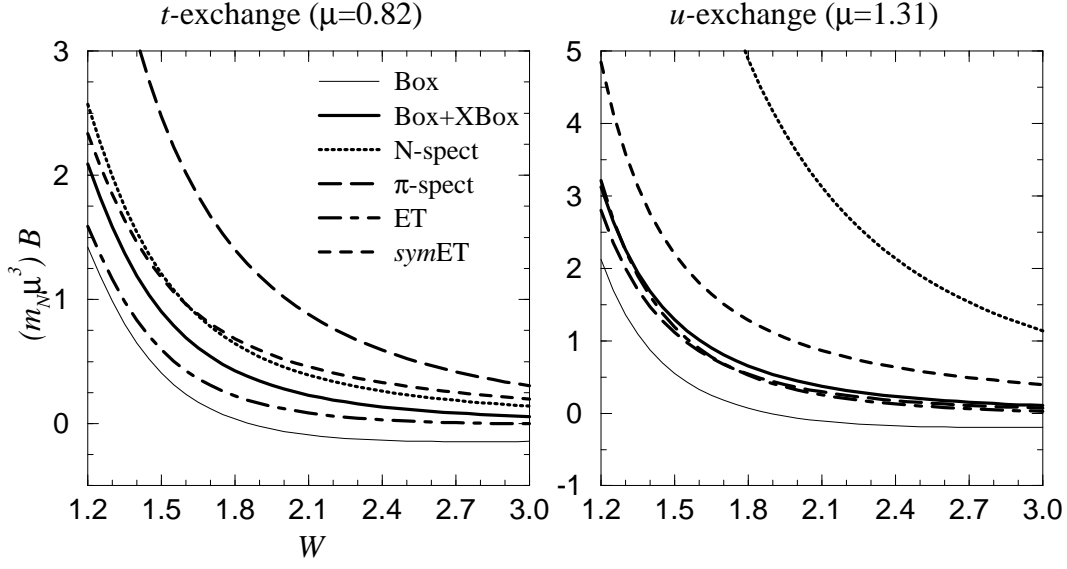}
\end{center}
\caption{ The same  as Fig. 6, but
for $m_N=1,\,m_\pi=0.15$ and $t=0$, as a function
of the energy $W$. Note that the mass of the exchanged particle
for the $t$- and $u$-exchange cases is not the same.}
\figlab{ex3}
\end{figure}

\vspace{3cm}

\epsfxsize=14.7cm
\begin{figure}[h]
\begin{center}
\epsffile{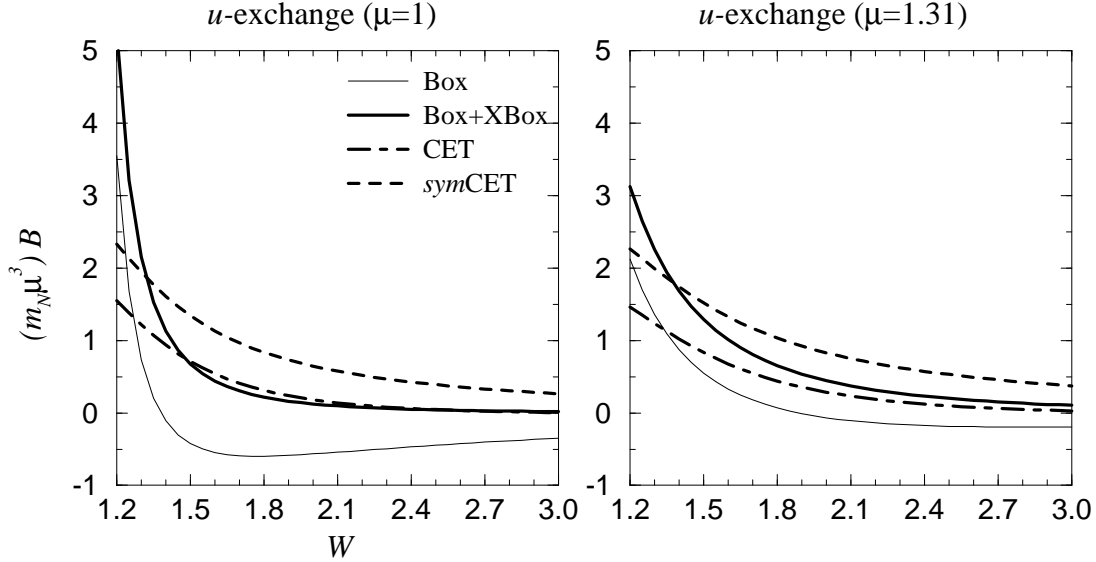}
\end{center}
\caption{ The QFT and CET  predictions
for $m_N=1,\,m_\pi=0.15$ and $t=0$. 
Exchange particle masses $\mu=1$ and $\mu=1.31$ have
been used for the left and right panels respectively.  }
\figlab{excet2}
\end{figure}

\newpage
\vspace{3cm}

\epsfxsize=15cm
\begin{figure}[h]
\begin{center}
\epsffile{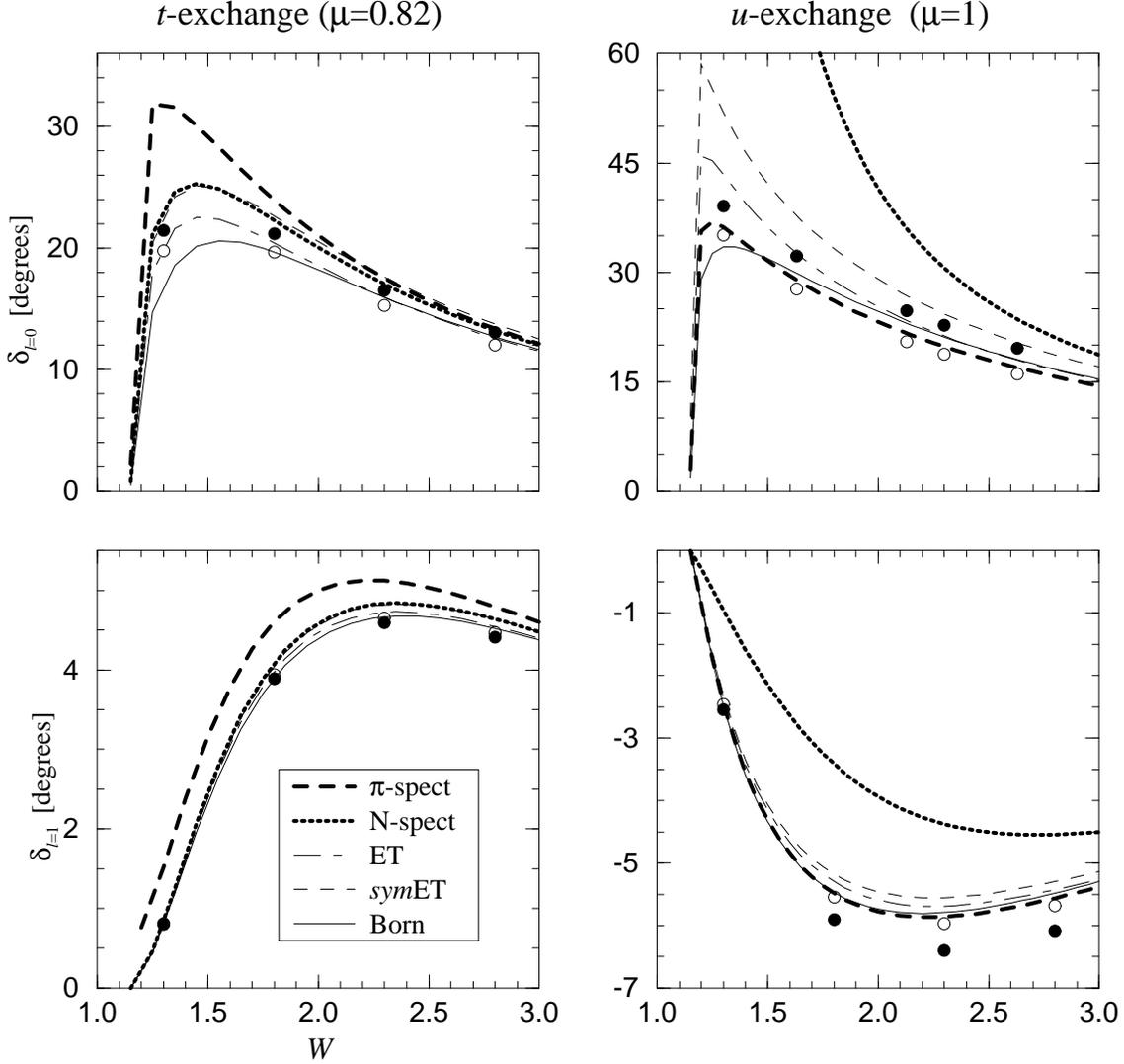}
\end{center}
\caption{S- and P-wave phase shift predictions of various QP
approximations with $m_N=1$, $m_\pi=0.15$, and $g^2/(4\pi)=2.0$. 
In the $u$-exchange S-wave case the nucleon spectator model 
predicts a phase shift which varies from $\pi$ to zero degrees,
i.e. supporting one bound state in this channel.
The open dots represent the results of the field-theoretic
box graph, the filled dots include the crossed-box
graph in addition.}
\figlab{phases}
\end{figure}

\newpage
\vspace{3cm}

\epsfxsize=15cm
\begin{figure}[h]
\begin{center}
\epsffile{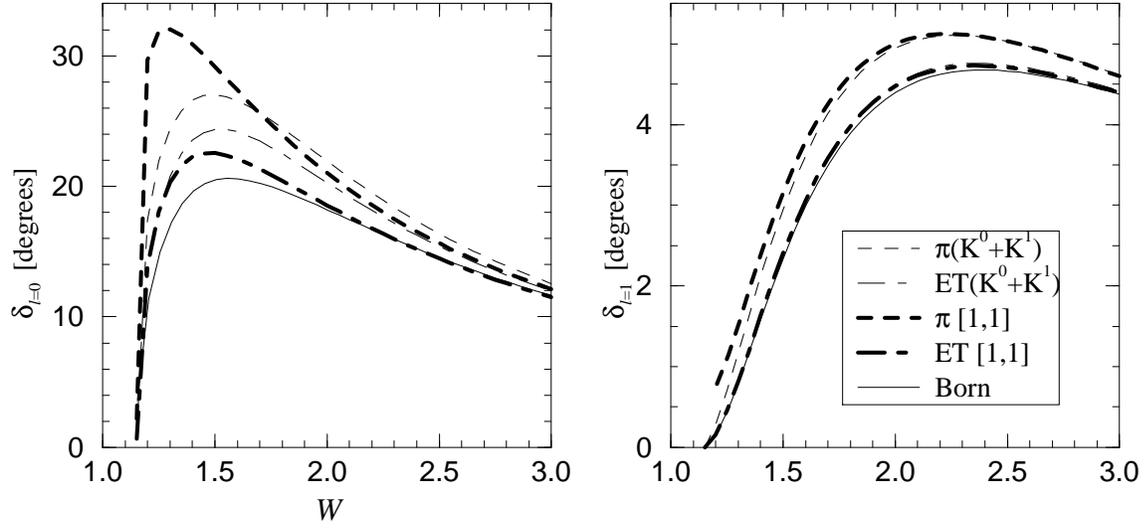}
\end{center}
\caption{Comparison of the phase shifts 
obtain from perturbation series  and the Pad\'e [1,1] approximant for
$t$-exchange potential.
The same set of parameters as in \Figref{phases} are used.
In the right panel the predictions of ET(K$^0$+K$^1$) coincide
with the ET $[1,1]$ Pad\'e approximant.}
\figlab{rescat}
\end{figure}

\end{document}